\documentclass[aps,prx,superscriptaddress,reprint,longbibliography,floatfix]{revtex4-2}
\usepackage{graphicx}
\usepackage{amsmath}
\usepackage{amssymb}
\usepackage{bm}
\usepackage{color}
\usepackage{float}
\usepackage[utf8]{inputenc}
\usepackage[T1]{fontenc}
\usepackage{newunicodechar}
\usepackage[colorlinks=true, allcolors=blue]{hyperref}
\usepackage{xcolor}
\usepackage{soul}
\setstcolor{red}
\usepackage[normalem]{ulem}

\begin{document}

\title{Extensive Spatio-temporal Chaos in Non-reciprocal Flocking}
\author{Chul-Ung Woo}
\affiliation{Department of Theoretical Physics and Center for Biophysics, Saarland University, Saarbr\"ucken, Germany}
\author{Jae Dong Noh}
\affiliation{Department of Theoretical Physics and Center for Biophysics, Saarland University, Saarbr\"ucken, Germany}
\affiliation{Department of Physics, University of Seoul, Seoul 02504, Korea}
\author{Heiko Rieger}
\affiliation{Department of Theoretical Physics and Center for Biophysics, Saarland University, Saarbr\"ucken, Germany}

\date{\today}

\begin{abstract}

Non-reciprocal interactions in active matter give rise to a multitude of fascinating phenomena among which are collective oscillatory states without intrinsic particle chirality and active turbulence. Here we show that in a paradigmatic model for non-reciprocal flocking, the two species Vicsek model, these two states coexist: chiral order for small flocks, and extensive spatiotemporal chaos for large flocks, both separated by a finite-wavelength instability whose scale is set by the rotation radius of the chiral orbits. For system sizes larger than this length scale extensive spatiotemporal chaos unfolds, as manifested by an extensive number of positive Lyapunov exponents as well as of  Floquet exponents, a finite correlation and chaotic length and a broad energy spectrum. Our results suggest that complex, turbulent behavior is a generic possibility in systems where particles or fields interact asymmetrically, and may have significant implications for understanding how non-reciprocal interactions could drive chaotic, fluid-like behavior in active matter.

\end{abstract}

\maketitle

\textit{Introduction}.
Turbulence in passive fluids has long served as a paradigmatic example of collective dynamics in spatially extended nonequilibrium systems~\cite{frisch1996turbulence,cross1993pattern}.
In active matter, related phenomena appear as \emph{active turbulence} in bacterial suspensions and active nematics~\cite{wensink2012meso,dunkel2013fluid,alert2022active,bratanov2015new}, where energy is injected at intrinsic small scales rather than transferred through an inertial cascade from the largest scales~\cite{alert2022active}.
From a broader nonlinear-dynamics perspective, turbulence is one manifestation of \emph{extensive spatio-temporal chaos}, which is a dynamical state of an 
extended system that can be decomposed into many weakly coupled chaotic subsystems, such that the number of positive Lyapunov exponents and the associated Kolmogorov-Sinai entropy scale extensively with system size~\cite{Ruelle1982large,xi2000extensive,tajima2002microextensive}.
Classical examples include Rayleigh-B\'enard convection~\cite{egolf2000mechanisms}, reaction-diffusion systems~\cite{kuramoto1978diffusion}, coupled oscillators~\cite{hu2000low}, and complex Ginzburg-Landau dynamics~\cite{aranson2002world}.
A common route to such behavior is a \emph{finite-wavelength instability} (FWLI)~\cite{cross1993pattern}: sufficiently small systems may sustain a regular pattern or a nearly homogeneous state, whereas beyond a characteristic size the dynamics crosses over to spatio-temporal chaos.
The Kuramoto-Sivashinsky equation provides a canonical example of such a system-size-controlled transition~\cite{Kuramoto-Book,papageorgiou1991route,wittenberg1999scale}.

Recent work has shown that in active matter \cite{marchetti2013hydrodynamics,bechinger2016active,rieger2020review,tevrugt2025metareview} non-reciprocal interactions, which violate action-reaction 
symmetry, can generate increasingly complex spatio-temporal dynamics.
In scalar active mixtures, it leads to moving and phase-coexisting structures~\cite{dinelli2023non} and to chaotic chasing bands in both particle-based and continuum models~\cite{duan2023dynamical,duan2025phase,brauns2024conserved,saha2025effervescence}.
In flocking mixtures, it likewise gives rise to spatially organized dynamical states, including synchronized bands, chase-and-rest dynamics~\cite{martin2025transition}, and species-segregated patterns under exclusive forcing~\cite{kreienkamp2024nonreciprocal} or weak non-reciprocity~\cite{myin2025flocking}.
More recently, non-reciprocal couplings have also been shown to produce turbulent regimes in active hydrodynamics~\cite{klamser2025directed,maji2026nonreciprocal}.
These studies make clear that non-reciprocity can drive active matter far beyond simple stationary states.
Whether non-reciprocal interactions can also invoke 
extensive spatio-temporal chaos in active matter systems other than
hydrodynamic turbulence remains elusive.

To shed light on this question we therefore consider a paradigmatic model
for flocking, the Vicsek model, with two non-reciprocally interacting species.
For this system, a phase with chiral order was
recently predicted
\cite{fruchart2021non,martin2025transition}, which is remarkable
because it implies the emergence of global rotations of the order parameter in a system of non-chiral particles, in contrast to the chiral systems studied in
\cite{liebchen2022chiral,mecke2024emergent,sumino2012large,giomi2013defect,wensink2012meso,dunkel2013fluid,han2020emergence,wioland2013confinement,lushi2014fluid}. {
As long as density fluctuations are neglected one expects 
quite generally the periodic motion of a macroscopic order parameter (in flocking systems the angle of the direction of motion) to be described by the KPZ equation
\cite{kardar1986dynamic,grinstein1993temporally,chate1995long,daviet2025kardar,maitra2025activity,chen2026only}, indicating the absence of true long-range order. 
} In this Letter we show for the non-reciprocal Vicsek model, which includes density fluctuations,
that chiral order survives only below a critical system size set by a FWLI.
Beyond this length scale, set by the rotation radius of the chiral orbits,
the dynamics exhibits extensive spatio-temporal chaos as we show by analyzing
the Boltzmann equation.

\begin{figure*}[t]
\centering
\includegraphics[width=\textwidth]{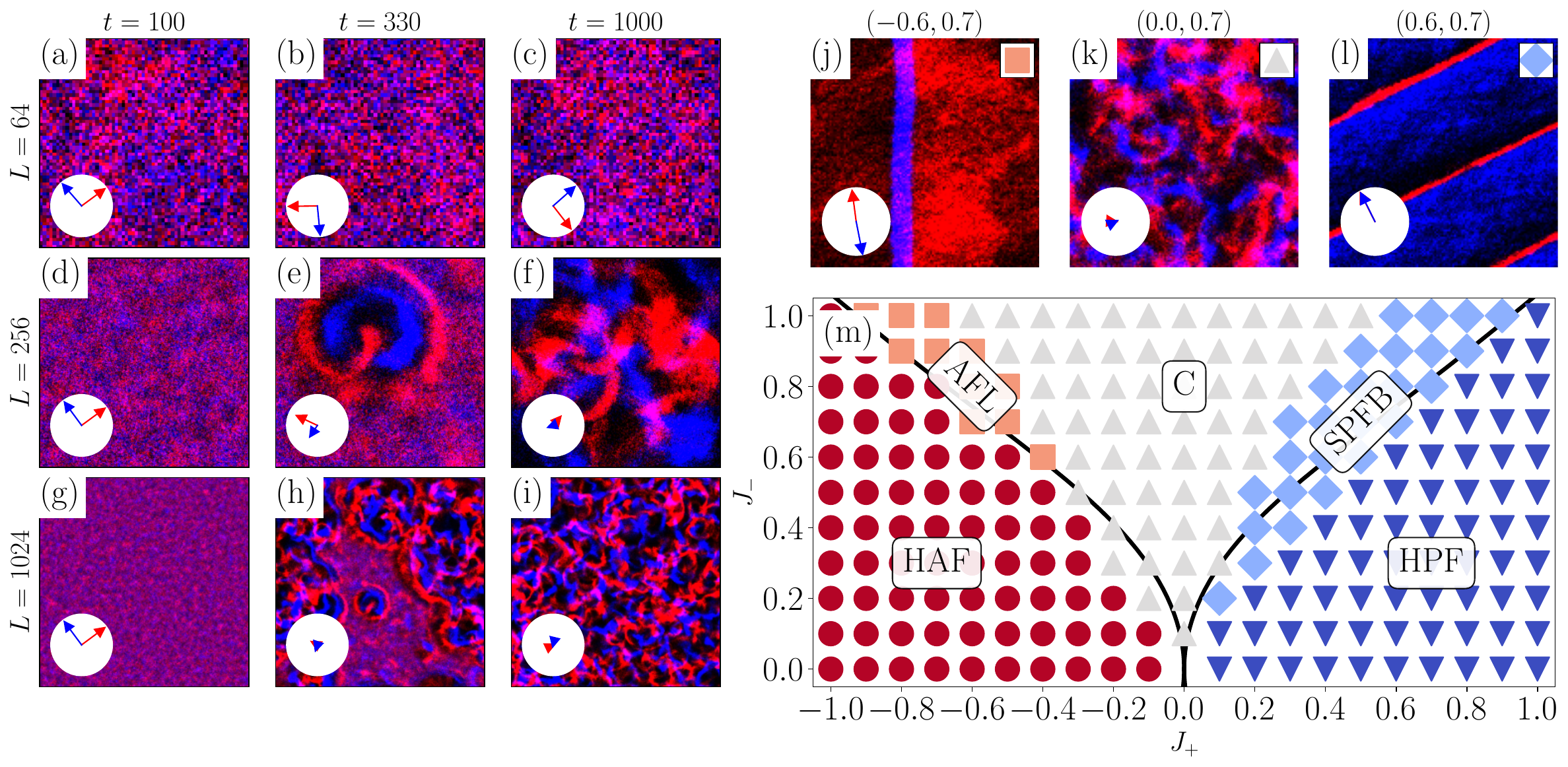}
\caption{Microscopic dynamics and phase behavior of the two-species non-reciprocal Vicsek model.
Species A is shown in red and species B in blue.
(a)--(i) Snapshots at $J_-=0.3$ for $L=64,256,1024$ (top to bottom) at $t=100,330,1000$ (left to right).
White insets indicate the instantaneous global polarizations of the two species.
(j)--(l) Representative steady-state snapshots at $J_-=0.7$, $J_+=-0.6,0,0.6$, and $L=256$.
(m) Microscopic phase diagram in the $(J_+,J_-)$ plane.
Circles denote homogeneous {antiparallel flocking (HAF)}, squares a {longitudinal antiparallel flocking lane (AFL)}, upward triangles the chaotic phase (C), diamonds a {species-separated parallel flocking band (SPFB)}, and downward triangles homogeneous {parallel flocking (HPF)}.
The black line indicates the exceptional transition predicted by the mean-field theory of Ref.~\cite{fruchart2021non}.}
\label{fig:fig1}
\end{figure*}

\textit{Model}.
We consider a two-species Vicsek-type model with non-reciprocal alignment in two dimensions.
Particle $i$ of species $s_i\in\{A,B\}$ has position ${\bm r}_i$ and heading $\theta_i$, and moves at constant speed $v_0$ in a periodic square of linear size $L$ according to
\begin{align}
\dot{\bm r}_i &= v_0 (\cos\theta_i,\sin\theta_i),\\
\dot{\theta}_i &= \sum_{\alpha\in\{A,B\}} \frac{J_{s_i\alpha}}{N_i^\alpha}
\sum_{j\in \mathcal N_i^\alpha}\sin(\theta_j-\theta_i)
+\sqrt{2T}\xi_i(t),
\label{eq:model}
\end{align}
where $\mathcal N_i^\alpha$ denotes the neighbors of species $\alpha$ within metric interaction range $R$ around $i$, $N_i^\alpha=|\mathcal N_i^\alpha|$, and the corresponding term is taken to vanish if $N_i^\alpha=0$.
The noise is Gaussian and white, with $\langle \xi_i(t)\xi_j(t')\rangle=\delta_{ij}\delta(t-t')$.
Unlike the agent-based dynamics used in Ref.~\cite{fruchart2021non}, each alignment contribution here is normalized by the instantaneous number of neighbors, $N_i^\alpha$.
Although this normalization can affect quantitative details, it does not alter the main qualitative conclusion~\cite{SM}.

We set $J_{AA}=J_{BB}=1$, while non-reciprocity is introduced through unequal interspecies couplings $J_{AB}\neq J_{BA}$.
It is convenient to parameterize them by $J_\pm=(J_{AB}\pm J_{BA})/2$, such that $J_-=0$ is the reciprocal limit and $J_+=0$ the purely antisymmetric line.
Unless stated otherwise, we use equal species densities $\rho_A=\rho_B=3$, interaction range $R=1$, self-propulsion speed $v_0=5$, noise strength $T=0.1$, and $J_+=0$.

\textit{Length-scale-dependent instability of the chiral state}.
In the regime of strong non-reciprocity, $J_- >J_+$, which we consider here,
the mean-field theory of Ref.~\cite{fruchart2021non} 
reported a spatially homogeneous time-periodic state
and predicted a stable phase with chiral long-range order.
We probe the stability of this homogeneous chiral~(HC) state using Monte Carlo simulations with an initial condition in which the global polarizations of the two species are offset by $\pi/2$. We find that the stability is strongly system-size dependent [Fig.~\ref{fig:fig1}(a)--(i)].
While the system of size $L=64$ maintains chiral order throughout the observation time window, chiral order disappears in larger system{s}: apparently triggered by local defects, the system evolves towards a strongly inhomogeneous state with pronounced species segregation. 
The decay of chiral order sets in earlier at systems with larger $L$.

The steady-state phase behavior at large scales is summarized in Fig.~\ref{fig:fig1}(m).
Along the cut $J_-=0.7$, the system passes through a sequence of distinct states as $|J_+|$ is reduced:
for large negative and positive $J_+$, it remains in the trivial homogeneous antiparallel and parallel flocking states, respectively, while at intermediate values it develops a longitudinal antiparallel flocking lane [Fig.~\ref{fig:fig1}(j)] for $J_+<0$ and a species-separated parallel flocking band [Fig.~\ref{fig:fig1}(l)] for $J_+>0$.
Near $J_+=0$, these ordered structures give way to a chaotic phase [Fig.~\ref{fig:fig1}(k)].
The black line in Fig.~\ref{fig:fig1}(m) indicates the exceptional transition predicted by the mean-field theory of Ref.~\cite{fruchart2021non}.
The main qualitative deviation from the phase diagram in 
\cite{fruchart2021non} occurs near the antisymmetric line $J_+=0$, where the mean-field HC state claimed in~\cite{fruchart2021non} is replaced by the chaotic phase.
In the following, we show that the decay of chiral order is controlled by a FWLI.

\begin{figure}
\centering
\includegraphics[width=\columnwidth]{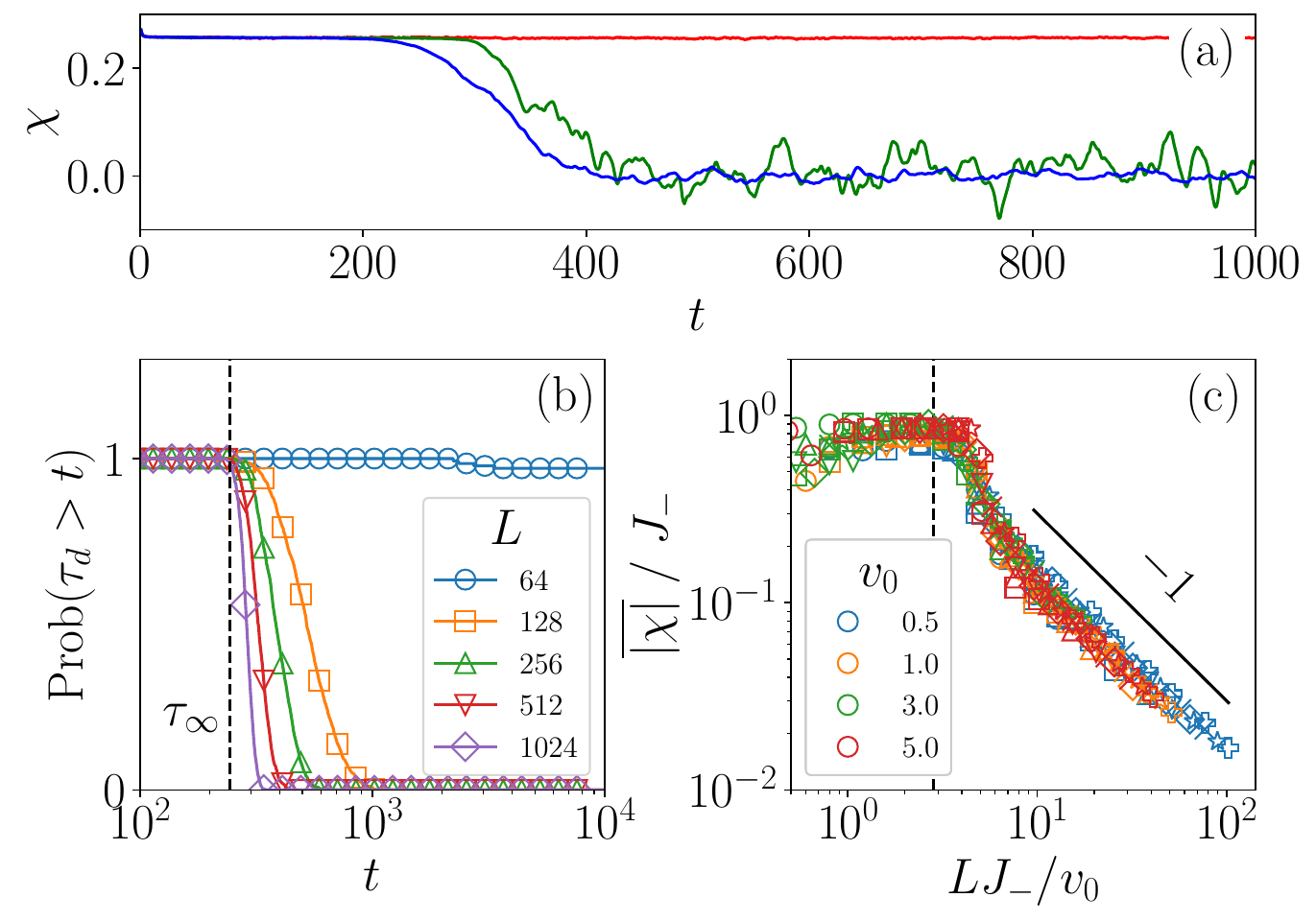}
\caption{(a) Time evolution of the global chirality $\chi(t)$ of Fig.~\ref{fig:fig1}(a)--(i); the three curves correspond to $L=64$~(red), $256$~(green), and $1024$~(blue).
(b) Survival probability $\mathrm{Prob}(\tau_d>t)$ at $J_-=0.2${, where $\tau_d$ denotes the onset time of the local breakdown of the prepared HC state}; the dashed line marks the decay time $\tau_\infty$ in the infinite size limit.
(c) Finite-size data collapse analysis of the scaled time-averaged chirality $\overline{|\chi|}/J_-$ against a scaling variable $LJ_-/v_0$ for different self-propulsion speeds $v_0$.
The vertical dashed line indicates the geometric crossover estimate, $2\sqrt{2}$. 
Different marker shapes indicate different values of $J_-$; the full marker legend is provided in \cite{SM}.
}
\label{fig:fig2}
\end{figure}

\textit{Global chirality and decay time}.
We define the global chirality $\chi$ as the particle-averaged deterministic part of the angular velocity, i.e., the right-hand side of Eq.~\eqref{eq:model} without the noise term.
Thus, $\chi$ measures the net collective rotation and vanishes in the absence of collective chiral motion.
As shown in Fig.~\ref{fig:fig2}(a), $\chi(t)$ remains finite in small systems, indicating a stable HC state over the observation window.
By contrast, in larger systems $\chi(t)$ decays to zero in a characteristic decay time, signaling the destabilization of the HC state.

To analyze the mechanism leading to the destruction of the chiral state we determine its decay time $\tau_d$. 
We define $\tau_d$ as {the onset time of the localized species-segregated breakdown visible in Fig.~1(e), using a criterion specified in~\cite{SM}.}
Figure~\ref{fig:fig2}(b) shows the survival probability $\mathrm{Prob}(\tau_d>t)$ for several system sizes at fixed $J_-=0.2$.
The survival probability converges to a finite value for $L=64$, indicating apparent long-lived stability of the chiral state, but decays to zero for larger values of $L$.
The survival probability curve {suggests convergence} to a step function sitting at finite $\tau_\infty$.
This behavior is incompatible with a picture of instability due to spatially independent Poisson nucleation events of defects, which in two dimensions would instead imply a characteristic decay time scaling as $L^{-2}$.
This differentiates the present mechanism from that reported in the active Ising model where an ordered phase is metastable due to spontaneously nucleated droplets~\cite{benvegnen2023metastability,woo2024motility}. 
The present mechanism is more consistent with a FWLI, which we will confirm below.

The long-time-averaged chirality obeys a simple finite-size scaling form.
Since the microscopic rotation rate is proportional to $J_-$, it is natural to write
$\chi(J_-;L,v_0)=J_- \chi_r(x)$ with $x=LJ_-/v_0\sim L/r_{\rm rot}$, where
$r_{\rm rot}\sim v_0/J_-$ is the rotation radius of a chiral orbit.
Figure~\ref{fig:fig2}(c) shows that the data collapse well when $\overline{|\chi|}/J_-$ is plotted against the scaling variable $LJ_-/v_0$.
The vertical dashed line indicates a crossover scale, which is estimated geometrically using an ideal circular orbit~\cite{SM}. The crossover scale estimated in this way will be called a geometric crossover estimate. 

For small $x$, $\chi_r(x)$ is approximately constant, corresponding to the regime in which the HC state remains stable.
For large $x$, the data decay approximately as $x^{-1}$, indicating that the global chirality vanishes in the thermodynamic limit.
This collapse identifies $r_{\rm rot}$ as the intrinsic length scale controlling the finite-size crossover.
When $L$ is not larger than this rotation-controlled scale, each particle of one species explores a substantial fraction of the orbit of the other species, so that the inter-species coupling becomes mean-field-like. 
As we show below, the wavelength selected by the kinetic instability is tied to this same intrinsic scale.

\begin{figure}
    \centering
    \includegraphics[width=0.99\linewidth]{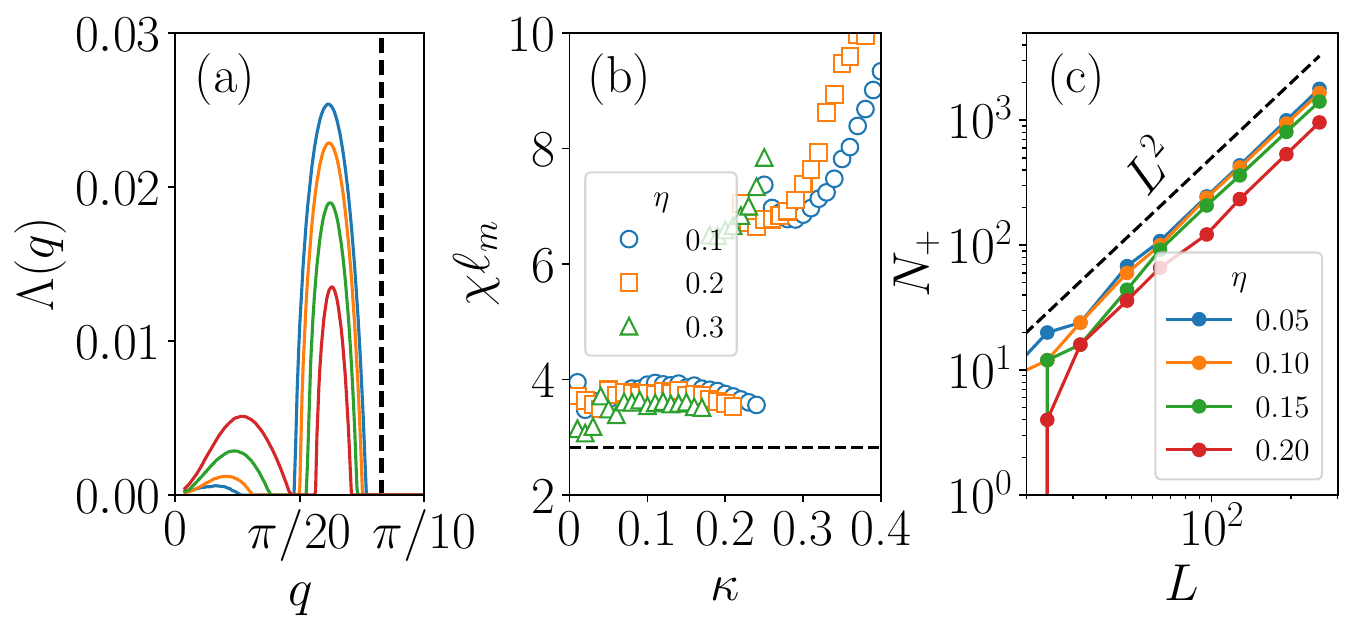}
    \caption{(a) Floquet exponents $\Lambda(q)$ as functions of wave number $q$ at several value{s} of $\eta$ at $\kappa=0.1$. 
    The colors are the same as in panel (c).
    (b) Rescaled wavelength $\chi \ell_m$ of the most unstable mode as a function of $\kappa$ for several value{s} of $\eta$.
    (c) Number of positive Floquet exponents, $N_{+}$, as a function of system size $L$.
    The dashed line is a guide to the eye proportional to $L^2$.
    }
    \label{fig:fig3}
\end{figure}

\begin{figure*}
    \centering
    \includegraphics[width=0.99\linewidth]{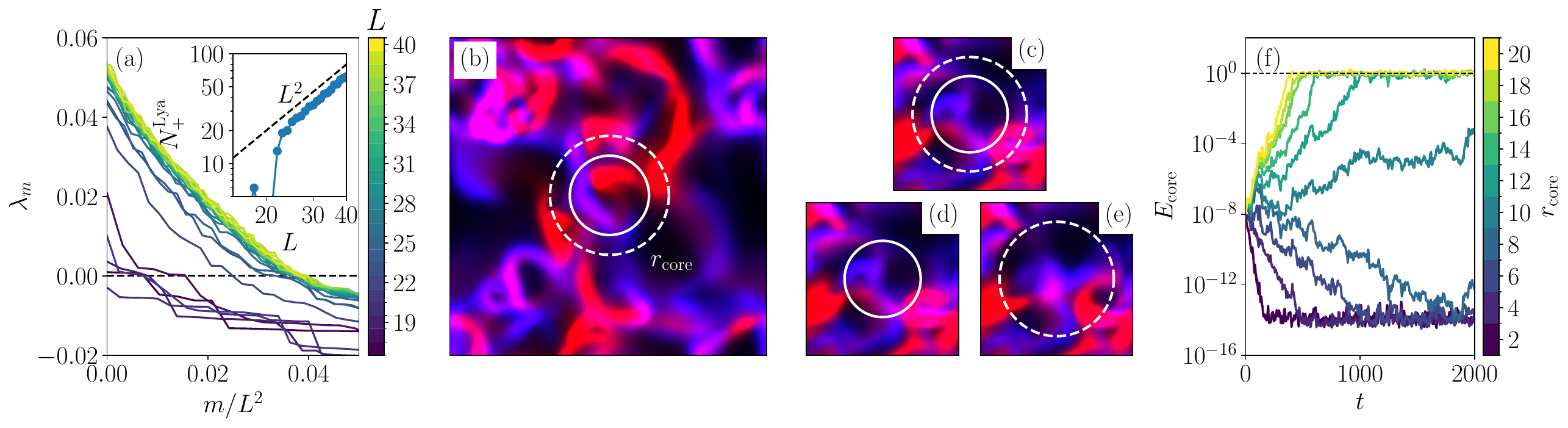}
    \caption{(a) {Lyapunov spectrum $\lambda_m$ plotted against the rescaled index $m/L^2$. The inset shows that $N_+^{\mathrm {Lya}}$, the number of positive Lyapunov exponents, scales as $L^2$.}
(b) A snapshot of a chaotic master state; the solid and dashed circles indicate two representative free core radii {$r_{\rm core}$}.
(c) Enlarged view of the master state in (b) after some time.
(d) and (e) show snapshots of slave states associated with the free cores with $r_{\rm core}=8$ and $12$, respectively.
(f) Mismatch $E_{\rm core}$ versus time for different $r_{\rm core}$. 
Slave states with small cores synchronize to the master state, whereas with large cores they evolve independently, revealing a finite chaotic length scale.
Parameters: $\kappa=0.2$, $\eta=0.1$. (b)-(f) $L=64$. 
}
    \label{fig:fig4}
\end{figure*}

\textit{Boltzmann kinetic description}.
To connect the microscopic dynamics to a continuum description, we introduce for each species $\alpha\in\{A,B\}$ a one-particle distribution $f^\alpha({\bm r},\theta,t)$.
Under the assumptions of dilute binary collisions and molecular chaos, in the spirit of standard Boltzmann descriptions of aligning active particles~\cite{Bertin2006Boltzmann,bertin2009hydrodynamic}, it obeys
\begin{equation}
\partial_t f^\alpha+v_0\hat{\bm e}(\theta)\cdot\nabla f^\alpha
=I_{\rm sd}[f^\alpha]+\sum_{\beta\in\{A,B\}}I_{\rm coll} [f^\alpha,f^\beta].
\label{eq:boltzmann}
\end{equation}
Here, $\hat{\bm{e}}(\theta) = (\cos\theta,\sin\theta)$,  $I_{\rm sd}$ describes angular self-diffusion, and $I_{\rm coll}$ binary interactions.

After nondimensionalizing space and time, we expand $f^\alpha({\bm r},\theta,t)$ in angular Fourier modes $f_k^\alpha({\bm r},t)$.
The zeroth and first modes, $f_0^\alpha$ and $f_1^\alpha$, are proportional to the density and the complex polarization field, respectively.
The resulting mode hierarchy reads
\begin{equation}
\begin{aligned}
\partial_t f_k^\alpha &
+\left(\partial_{z^*}f_{k-1}^\alpha+ \partial_z f_{k+1}^\alpha\right)
\\
&= -(1-P_k)f_k^\alpha
+\sum_{\beta}\sum_{l=-\infty}^{\infty}
\kappa^{\alpha\beta}J_{k,l}^{\alpha\beta} f_{k-l}^\alpha f_l^\beta ,
\end{aligned}
\label{eq:BE_mode_hierarchy}
\end{equation}
where $z=x+iy$, $\partial_z=(\partial_x-i\partial_y)/2$, and $\partial_{z^*}=(\partial_x+i\partial_y)/2$.
Here $P_k$ is the $k$th Fourier coefficient of the self-diffusion kernel, $J_{k,l}^{\alpha\beta}$ encodes the nature of the interactions (aligning or anti-aligning), and the coefficients $\kappa^{\alpha\beta}$ set the interaction strength, see \cite{SM}.
Throughout this work, we fix $\kappa^{AA}=\kappa^{BB}=0.5$ and vary $\kappa^{AB}=\kappa^{BA}=\kappa$.
With this choice, the model corresponds to the microscopic model with $J_+=0$ and $J_->0$.
To obtain a closed kinetic description, we truncate the angular Fourier hierarchy at $k=4$.

\textit{Floquet stability analysis}.
We found that the Boltzmann equation limited to a homogeneous manifold allows a HC solution in which the polarizations perform limit cycle motion of period $\mathcal{T}$ and chirality $\chi=2\pi/\mathcal{T}$.
We perform a quantitative analysis of the stability of this state
against a position-dependent random perturbation using the Floquet method~\cite{barkley1996threedimensional}. 

It is convenient to work in the spatial Fourier space labeled by a wave vector $\bm{q}$. 
Taking the HC solution, computed numerically, as a reference state, we initialize random perturbations for all Fourier modes with nonzero $\bm{q}$. 
This perturbed state is evolved over one period $\mathcal{T}$, the growth-rate of each mode at ${\bm q}$ is measured, and the amplitudes of perturbed modes are rescaled back to their initial values. 
Repeating this procedure iteratively, we can obtain a local Floquet exponent $\Lambda(\bm{q})$. 

Figure~\ref{fig:fig3}(a) shows $\Lambda(q)$, radially averaged $\Lambda(\bm{q})$ over all ${\bm q}$ with given $q=|{\bm q}|$, for several values of $\eta$ at fixed $\kappa = 0.1$.
In all cases considered, the dominant peak lies at finite $q$, showing that the HC motion is destabilized by a finite-wavelength mode.
The corresponding wavelength $\ell_m=2\pi/q_m$, determined by the wavenumber $q_m$ of the maximally unstable mode, is shown in Fig.~\ref{fig:fig3}(b) through the rescaled quantity $\chi \ell_m$.
In the weak-coupling regime, $\chi \ell_m$ remains approximately constant. This behavior is on par with the crossover in the microscopic model occurring at $L \sim r_{\rm rot}$.
We note that the plateau value of $\chi l_m$ is close to the geometric crossover estimate indicated by the horizontal dashed line.
Therefore we conclude that the FWLI of the Boltzmann equation is related to the  finite-size destabilization of the HC state in the particle model.
Deviations from this weak-coupling correspondence at larger $\kappa$ and $\eta$ are discussed in \cite{SM}.

Figure~\ref{fig:fig3}(c) shows the number of unstable modes $N_+$, defined as the number of Fourier modes with positive Floquet exponent.
This number grows linearly with the system area, $N_+\sim L^2$, 
meaning that it is extensive. 

\textit{Evidence for extensive spatio-temporal chaos (ESTC).}
{We proceed further beyond the linear Floquet stability analysis by calculating the Lyapunov exponents characterizing the exponential rate of separation of infinitesimally close trajectories in the steady state~(see~\cite{SM} for the numerical method). Figure~\ref{fig:fig4}(a) shows the Lyapunov exponents $\lambda_m$, sorted in descending order, in the chaotic regime. The largest Lyapunov exponent is positive. Moreover, the number of positive Lyapunov exponents $N_+^{\rm Lya}$ scales extensively with the system size as $L^2$, which is a hallmark of ESTC~\cite{xi2000extensive,tajima2002microextensive}. In~\cite{SM}, we show that the Kolmogorov-Sinai entropy is also extensive, the velocity-velocity correlations are short-ranged, and the energy spectrum is broad, which all point towards ESTC. }

{Complementary evidence} for ESTC comes from the observation that the chaotic state can be decomposed into weakly coupled chaotic areas of the size of the instability length-scale that we identified above.
For this purpose we implemented a master-slave protocol as described in the context of synchronization of chaotic PDEs and turbulence~\cite{kocarev1997synchronizing,lalescu2013synchronization,karimi2012length}.
As in the set up for the Lyapunov exponent, we prepare unperturbed and perturbed states, which are referred to as master and slave states, respectively.
We introduce a circular region~({\it free core}) of radius $r_{\rm core}$ (see Fig.~\ref{fig:fig4}(b)) and impose an additional uni-directional coupling:
the slave state evolves independently of the master state inside the free core, but is attracted toward the master state outside the free core.
If $r_{\rm core}$ is smaller than the chaotic length scale, the slave state within the free core gets enslaved by the field outside the free core, i.e., the master state. 
Otherwise, the slave state develops its own chaotic evolution.
Figures~\ref{fig:fig4}(d) and \ref{fig:fig4}(e) illustrate these two cases for a small and a large $r_{\rm core}$, respectively.
This behavior is quantified by a mismatch $E_{\rm core}(t)$, defined as a relative deviation between the master and slave states within the free core~(see \cite{SM}), shown in Fig.~\ref{fig:fig4}(f): 
for small cores the mismatch decays to zero, whereas for sufficiently large cores it remains finite at long times.
The crossover occurring at $r_{\rm core}\sim 10$ identifies a finite chaotic length scale, beyond which different regions behave as independent chaotic subsystems.
The corresponding diameter agrees with the system size $L\simeq 22$ above which $N_+^{\rm Lya}$ exhibits extensive scaling. 
Together with the diagnostics above, this consistency indicates that the large-scale kinetic state exhibits ESTC.

\textit{Discussion}.
We have shown that the homogeneous chiral state of the non-reciprocal Vicsek model is not the asymptotic large-scale state in two dimensions \footnote{
Notably, non-reciprocity due to small symmetric phase-shifts in the alignment
interactions between multiple species does exhibit quasi-long-range chiral order \cite{woo2025collective, woo2025nonreciprocal}}.
Instead, it exists only below a characteristic scale set by the radius of the chiral orbit, and is destabilized at larger scales due to a finite-wavelength instability. 
This size-dependence has to be distinguished from finite-size effects at conventional phase transitions which just smoothen the critical singularity
\cite{BinderFSS} and also from a crossover from chiral order to 
KPZ-behavior in chiral Malthusian flocks~\cite{chen2026only}. 
{
One reason why a KPZ behavior for periodic motion of a macroscopic order parameter 
\cite{kardar1986dynamic,grinstein1993temporally,chate1995long,daviet2025kardar,maitra2025activity,chen2026only} is not observable in the non-reciprocal Vicsek model might be the relevance of density fluctuations: 
in Fig. 1(e) and (h), as well as in movies 2 and 3 in~\cite{SM}, one sees that within the chiral ordered 
state first localized droplets of species-segregated density fluctuation emerge and then they expand and destroy chiral order.}
The FWLI is also different from metastability due to nucleation of local defects \cite{benvegnen2023metastability,woo2024motility}.
The system size acts here like a parameter triggering the transition from an ordered to a chaotic state, like frequently in turbulence and 
specifically in the Kuramoto-Sivashinsky equation 
\cite{Kuramoto-Book,cross1993pattern}.

In the non-reciprocal Vicsek model the transition is accompanied by the loss of global chirality and the emergence of extensive spatio-temporal chaos. In connection with recent works on non-reciprocal turbulence, notably in binary-fluid systems~\cite{klamser2025directed,maji2026nonreciprocal}, our results suggest that complex, turbulent behavior is a generic possibility in systems where particles or fields interact asymmetrically.
This may have significant implications for understanding how non-reciprocal interactions could drive chaotic, fluid-like behavior in active matter. 
It also provides a new mechanism for turbulence that does not require traditional geometric or external triggers.

\begin{acknowledgments}
This work was supported by the National Research Foundation of Korea~(NRF) grant funded by the Korea government~(MSIT) (Grant No. RS-2024-00348526) and 
in part by grant NSF PHY-2309135 to the Kavli Institute for Theoretical 
Physics (KITP). 
We acknowledge the Urban Big Data and AI Institute of the University of Seoul supercomputing resources made available for conducting the research reported in this paper.
\end{acknowledgments}

{{\it Note added in proof:} After the submission of this work, we became aware of a study of the discrete non-reciprocal Vicsek model reporting the destruction of chiral order by the nucleation of defects \cite{myin2026breakdown}.}

\end{document}